# The global distribution of active Ionian Volcanoes and implications for tidal heating models

Julie A. Rathbun - Planetary Science Institute
Rosaly M. C. Lopes – Jet Propulsion Laboratory, California Institute of Technology
John R. Spencer – Southwest Research Institution

**Abstract**
　　Tidal heating is the major source of heat in the outer Solar System.  Because of its strong tidal interaction with Jupiter and the other Galilean Satellites, Io is incredibly volcanically active.  We use the directly measured volcanic activity level of Io's volcanoes as a proxy for surface heat flow and compare to tidal heating model predictions.  Volcanic activity is a better proxy for heat flow than simply the locations of volcanic constructs. We determine the volcanic activity level using three data sets: Galileo PPR, Galileo NIMS, and New Horizons LEISA.  We also present a systematic reanalysis of the Galileo NIMS observations to determine the 3.5 $\mu$m brightness of 51 active volcanoes.
　　We find that potential differences in volcanic style between high and low latitudes make high latitude observations unreliable in distinguishing between tidal heating models. Observations of Io's polar areas, such as those by JUNO, are necessary to unambiguously understand Io's heat flow.  However, all three of the data sets examined show a relative dearth of volcanic brightness near 180 W (anti-Jovian point) and the equator and the only data set with good observations of the sub-Jovian point (LEISA) also shows a lack of volcanic brightness in that region.  These observations are more consistent with the mantle-heating model than the asthenospheric-heating model.  Furthermore, all three of the data sets are consistent with four-fold symmetry in longitude and peak heat flow at mid-latitudes, which best matches with the combined heating case of Tackley et al. (2001).

**Introduction**
　　Tidal heating is the major contributor to active surface geology in the outer Solar System, and yet its mechanism is not completely understood. Io's volcanoes are the clearest signatures of the tidal heating process, thus measurements of Io's total heat output and how it varies in space are useful constraints on tidal heating models.  The global distribution of surface heat flow is likely to be influenced by whether the tidal heating occurs primarily in Io's asthenosphere or mantle (Segatz et al., 1988; Gaskell et al., 1988; Ross et al., 1990; Tackley, et al., 2001; Hamilton et al., 2013).  The major difference is that asthenospheric heating predicts greater heat flow at mid-latitudes while mantle heating predicts higher heat flow at the poles.
　　Lopes-Gautier et al. (1999) and Hamilton et al. (2013) assumed that hotspots represent the major pathway of magma to Io's surface and therefore the spatial distribution of hotspots reflects the distribution of total heat flow.  Lopes-Gautier et al. (1999) used early data from the Galileo Near Infrared Mapping Spectrometer (NIMS) to find that the global distribution of major hotspots was more consistent with the asthenosphere model than with the deep mantle model.  By comparing the locations of hotspots to the heat flow patterns predicted by the two end-member models, Ross et al. (1990) found that the observed uniform hotspot distribution was more consistent with the asthenospheric model, although specific features in the heat flow pattern were not matched by the hotspots. Lopes-Gautier et al. (1999) also built on earlier work by Gaskell et al. (1988) by comparing the locations of hotspots to the large-scale topography. Gaskell et al. (1988) proposed that the large-scale topography might reflect isostatic compensation due to upwellings and thus that more hotspots would be expected at higher elevations. However, neither McEwen (1995) nor Lopes-Gautier et al. (1999) found any correlation between locations of hotspots and topography.

Kirchoff et al. (2011) looked at the global distribution of both Io's volcanic centers and mountains. They found that the volcanic centers have a very strong degree-2 distribution. That is, that the volcanic centers have two peaks and two minima in their geographic distribution. The peaks were located near the equator at ~170 W and 345 W. They found this most consistent with asthenospheric heating of Io, though with some rotation of the pattern, perhaps due to nonsynchronous rotation of Io's crust. Kirchoff et al. (2011) also found that mountains have a asymmetric degree-2 distribution with the major peak at 25 N, 65 W and the minor peak at 20 S, 265 W, and a global anticorrelation with volcanic centers.

Hamilton et al. (2013) used nearest neighbor analysis to analyze the distribution of hotspots, paterae, and paterae floor units from different available data sets and maps. They found that globally, hemispherically, and within the north and south polar regions, hotspots are distributed randomly and, therefore, thermal anomalies form independently of each other. This is consistent with the strong degree-2 signal found by Kirchoff, et al. (2011) because Hamilton et al. (2013) found randomness between pairs of features, and randomly spaced pairs of points can also be organized into larger clusters consisting of more than two members. Hamilton et al. (2013) further found that hotspots are more uniformly distributed in the near-equatorial region, which suggests that in the more populated near-equatorial region hotspots may tend to avoid each other to maximize the utilization of the available magma. They also found that paterae, analogous to terrestrial volcanic systems, also tend to be uniformly distributed, but everywhere, which suggests that vigorous mantle convection may help to smooth out the overall distribution of heat on Io. Lastly, they found that patera floor units tend to be clustered over multiple scales, which indicates that most volcanic systems are active long enough to feed multiple eruptive units. This pattern of clustering among patera floor units is consistent with the findings of Kirchoff et al. (2011) and demonstrates that the distribution of volcanic features on Io is complex with regional variations that imply spatial variability in Io volcanic processes.

These previous studies of the global distribution of hotspots only used a subset of the data that are available now, and they did not take into account the different activity levels of the hotspots (for example, Loki is significantly more powerful than any other hot spot on Io). While Veeder et al. (2009, 2011, 2012) included activity level in their analyses of the global variations in Ionian heat flow their analysis was model-dependent. They analyzed the level of activity at 240 different presumed thermal sources and calculated Io's thermal emission by latitude and longitude. While their calculated power outputs include the low temperature contribution that dominates Io's heat flow, most of the powers are model dependent and based on several assumptions. In many cases, they assume a temperature based on similar measured thermal areas and in all cases they assume the entire dark feature is warm which, according to analysis of high spatial resolution data from Galileo (Lopes et al., 2004) was not always the case. They found peaks in the thermal emission at 120 W and 330 W and a dearth at high latitudes. Davies et al. (2015) used the Veeder et al. (2009, 2011, 2012) data to create a map of volcanic heat flow. They found relatively low heat flow at the sub- and anti-Jovian points and that Io's anomalously warm poles could result from deep mantle heating. de Kleer and de Pater (2016) also considered activity level, as measured directly by ground-based adaptive optics systems between 2013 and 2015. They found that the brightest eruptions are more clustered and occur at higher latitudes.

Here, we use the power output of volcanoes as derived directly from measurements from three different instruments and compare to tidal heating models. We make the assumption that the spatial distribution of volcanic output, not simply volcanic constructs, reflects the distribution of total heat flow. The spatial distribution of volcanic constructs depends not only on how quickly they form (plausibly a proxy for heat flow), but how quickly they are destroyed. If the destruction processes are volcanic (flooding with lava or burial by pyroclastics), volcanic features could be destroyed more rapidly in more active regions. Less active regions might then have just as many volcanic features, but they would be older on average. Thus, using active volcanic output is a better proxy for heat flow than volcanic constructs. Furthermore, because we use power outputs and brightnesses derived directly from observations, there are no model assumptions used in their derivation.

**Available hotspot data sets**

Observations of hotspots on Io are available from several spacecraft. The Galileo spacecraft carried three remote sensing instruments useful for measuring the magnitude of volcanic activity. The Solid-State Imager (SSI) obtained high resolution, $0.38 - 1.04\ \mu$m images and was able to image endogenic activity at night, in eclipse, and, for very hot volcanoes, in the daytime (McEwen et al., 1997; McEwen et al., 1998). Deriving quantitative measurements of hot spot activity from the daytime images is complicated and made particularly difficult by pixel saturation and bleeding (Milazzo et al., 2005), so we do not use SSI data in the analysis presented here. Furthermore, it shows only the hottest activity and is less likely to be representative of total heat flow than longer wavelengths.

The Galileo Photo-Polarimeter Radiometer (PPR) instrument, when used in radiometry mode, measures the brightness temperature of the surface at long infrared wavelengths of $17 - 100\ \mu$m (Russell et al., 1992). Rathbun et al. (2004) calculated total hotspot power output from PPR datasets obtained at different times. They subtracted a background power calculated based on the average temperature of points near the hot spot. In cases where the same hotspot was observed more than once, they calculated the ratio between the maximum and minimum power. They found that the most dramatic changes occurred at Dazhbog, Hephestus, and Ot, which varied by a factor of several over a period of less than two years. Because the study was limited to nighttime data only, their results only cover longitudes of 160 – 360 W.

The Galileo NIMS instrument (Carlson et al., 1992) obtained spectral image cubes with as many as 408 and as few as 12 spectral channels between 0.7 and $5.2\ \mu$m (Lopes et al., 2004). For data obtained at night or in eclipse, simply fitting one or more blackbody curves to the data yields quantitative information about the volcano's activity. For daytime data, only the longer wavelengths are fitted, to minimize the inexact subtraction of reflected sunlight. Lopes-Gautier et al. (1999) quotes $4.7\text{-}\mu$m brightnesses to compare volcanic output. In the next section, we present our own systematic analysis of NIMS data cubes to extract $3.5\ \mu$m brightnesses of every observed active volcanic center.

The New Horizons spacecraft also carried three remote sensing instruments useful for measuring thermal volcanic output (Reuter et al., 2008; Cheng et al., 2008). The Long-Range Reconnaissance Imager (LORRI) obtains high spatial resolution visible light images in a single broad wavelength band, the Multicolor Visible Imaging Camera (MVIC) obtains lower resolution images simultaneously in four wavelength bands, and the Linear Etalon Imaging

Spectral Array (LEISA) obtains spectral image cubes at wavelengths from 1.25 to 2.5 $\mu$m. Spencer et al. (2007) presented preliminary measurements of volcanic output at Tvashtar and East Girru based on blackbody fits to LEISA spectra. Rathbun et al. (2014) presented a thorough analysis of LORRI and MVIC observations of active Ionian volcanoes. Tsang et al. (2014) determined the location of every active volcano observed by LEISA and single temperature blackbody fits to those with enough signal-to-noise for a reasonable fit. They found 52 observations of 37 distinct active volcanoes, and were able to fit a blackbody to 26 of those observations. However, because of the relatively short wavelengths observed, these measurements are not the best source for comparison to heat flow models. Furthermore, they are at shorter wavelengths than NIMS and PPR, but the longest available from New Horizons, so we use those in this study.

    The available datasets are complementary in time, space, and wavelength coverage. Here, we chose three of the above-discussed data sets (PPR, NIMS, and LEISA) and use the derived volcano brightnesses to compare to tidal heating models. All of the data sets are obtained over geologically short periods of time, so any changes are not due to long term variations in heat flow and we can ignore changes in volcano strength by averaging the observations from a single instrument for any given volcano. In all cases, we are using either direct observations or blackbody fits, so there are no major modeling assumptions used.

    Galileo PPR obtained the longest wavelength data, therefore the data that most closely tracks heat flow. Rathbun et al. (2004) derived total power output from each observed volcano, so we can use these powers directly to compare to heat flow models. Unfortunately, PPR did not obtain global coverage. However, since the coverage was relatively complete in latitude, these data can be used to discriminate between heat flow models. Based on the extent of the mosaic image used by Rathbun et al. (2004) to detect volcanoes, we consider only latitudes below 80° and longitudes between 160° and 330° to be well covered by this instrument. In other areas, it is possible that active volcanoes exist but were not observed, because they were beyond the scope of that dataset.

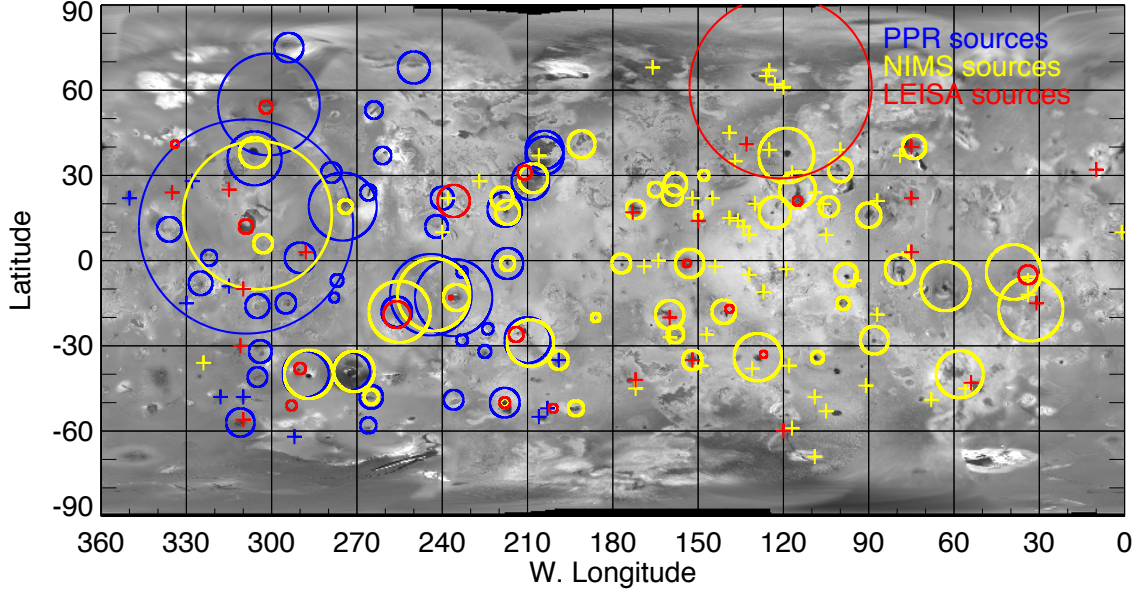

**Figure 1:** Location and relative strength of all hotspots used in this study. Each circle is centered on the latitude and longitude of the observed hotspot. For the PPR data, the area of the circle is proportional to the measured power output from the hotspot (Rathbun et al., 2004). For the NIMS data, the area of the circle is proportional to the average measured 3.5 micron brightness. For the LEISA data, the area of the circle is proportional to the 2.2 micron brightness reported in Tsang et al. (2014). For all data sets, plus signs indicate a hotspot that was detected in the data, but the data were not good enough to obtain a measurement of its relative brightness. Note the scarcity of volcanoes near the anti-Jovian point (180 W longitude, 0 latitude).

Tsang et al. (2014) published derived 2.2 $\mu$m brightnesses for every volcano for which they were able to fit a blackbody spectrum to the LEISA data. Short-wavelength output of active volcanoes is more variable than the long-wavelength output (Davies et al., 2010), so these measurements may not be indicative of the long-lived level of activity of these volcanoes. LEISA's advantage is that it obtained global coverage of Io, with spatial resolutions better than 250 km/pixel at the equator for all longitudes. While this is the most spatially complete data set obtained with a single instrument, it favored the southern hemisphere due to New Horizon's subsolar latitude of 2.1 - 8.7 S during the observations. Correcting for emission angle by dividing by the cosine of the emission angle, the spatial resolution in the north-south direction is worse than 300 km/pixel poleward of 70 S and 50 N. We consider only latitudes below these extremes to be well covered by this instrument. Beyond these latitudes, it is possible that active volcanoes exist but were not observed.

We reanalyzed Galileo NIMS observations (see next section) and determined the 3.5-$\mu$m brightness of 51 active volcanic features. All of the 51 features were observed multiple times, so an average value is used in our analysis. NIMS observed at longer wavelengths than LEISA but shorter wavelengths than PPR, so it is also missing some of the long wavelength component of the power output, but not as much as LEISA. Furthermore, while all of the PPR and LEISA data used were obtained at night, to get global NIMS coverage, daytime images were also analyzed.

However, the method used for extracting emission spectra (discussed below) resulted in very noisy spectra for daytime hotspots and large uncertainties in derived quantities.

Figure 1 shows the locations and relative strength of active volcanoes measured by the three instruments discussed above. Each circle is centered on the location of the observed volcano and area of the circle is proportional to either the measured total power (PPR) or the measured brightness (3.5 $\mu$m for NIMS, 2.2 $\mu$m for LEISA). All three instruments were able to detect dimmer hotspots for which the power and brightness could not be determined; these are indicated in figure 1 as plus signs. The observational biases in the PPR data are obvious, with no volcanoes observed on the leading hemisphere. Note the scarcity of volcanoes at the anti-Jovian point (180 W), which is located in an area well covered by all three data sets.

### NIMS data analysis

We used data from the Galileo Near Infrared Mapping Spectrometer (NIMS) to systematically determine the brightness for each active volcano in each image. To do this, we used the NIMS cubes (Lopes et al., 2001) and began by converting the measured brightnesses to units that more easily allow us to calculate the total brightness for the observed active volcanic hotspots. Note that this is different from the methodology of Veeder et al. (2015) in that they used both tubes (original data) and cubes (spatially resampled data) in their analysis. Furthermore, we fit a single temperature blackbody to each spectrum while they used both a single temperature and a combination of two-temperature components to fit the nighttime data. For the daytime data, they assume an area based on the observed visible wavelength area of each dark feature and use a single brightness in the 4.7-5 micron range to derive a temperature for that area.

Radiances in NIMS cubes are given as power per unit area per $\mu$m per steradian in SI units (W/m$^2$/str/μm). We first determine the slant distance ($D$) of the spacecraft to the imaged surface for the observation, using the method outlined in the NIMS documentation on the PDS section 6D. The size of each pixel in the observation is then given by

$$A = (D*r)^2 / \cos(\varepsilon) \qquad (1.1)$$

where $r$ is the angular resolution of the NIMS pixel and $\varepsilon$ is the emission angle. By multiplying this area by the radiance we determine the total brightness from that location as measured by the instrument, in units of W / str / $\mu$m. At this point our processes for analyzing nighttime and daytime data diverge. It is easier to determine endogenic heating from nighttime data because all of the heating observed is endogenic.

### Nighttime Data

For each image, we interactively mark each hotspot. We determine which image of the image cube is best to use for noticing hotspots by comparing images at wavelengths longer than 3 μm and an integrated image obtained by totaling the area under the spectrum at wavelengths larger than 4.5 μm. Once the best image is selected, each hotspot in the image is selected. For each spot chosen, we sum the power from the 13 pixels closest to the chosen location. This approximate circle of 13 pixels enables us to account for brightness that leaked to adjacent pixels during the tube to cube conversion process without including too much flux from adjacent hotspots. For each hotspot, we note its location as well as total emission spectrum.

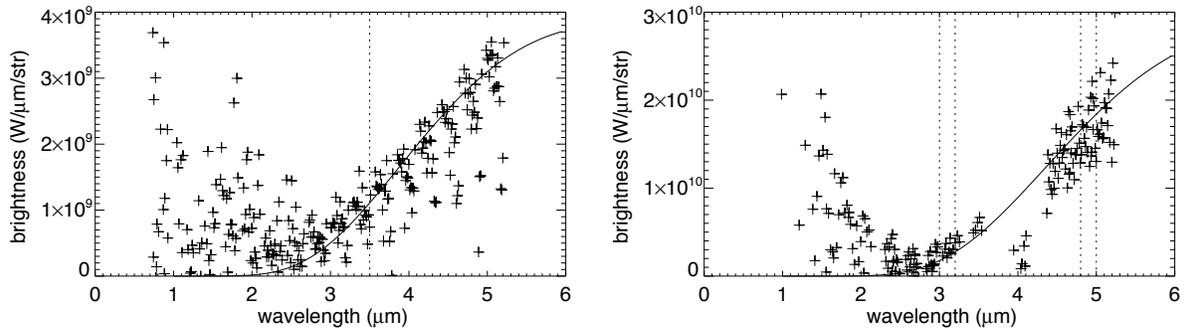

**Figure 2: Example data and single temperature blackbody fits to nighttime data (left) and daytime data (right). Data on the left is the total brightness of a hotspot located at 0 N 78 W in observation G1INNSPEC_01A, likely Hi'iaka. The solid line is a blackbody curve for a temperature of 439 K and an area of 57 km$^2$. The vertical line indicates the lowest wavelength used for the fit. Data on the right is the total brightness of a hotspot located at 18 N 174 W in G2INHRSPEC01A, likely Pillan. The solid line is a blackbody curve for a temperature of 392 K and an area of 751 km$^2$. The vertical lines indicate the two narrow wavelength bands that were used for the fit.**

We fit a single temperature blackbody separately to each of the positions of the NIMS grating (Carlson et al., 1992), since small motions of the target can occur between grating positions. We only fit wavelengths larger than 3.5 µm because lower wavelengths have particularly low signal to noise (Stephan, et al., 2008) and do not see a large contribution from thermal radiation at these temperatures. Moreover, the NIMS Detectors 1 and 2 (corresponding to the shorter wavelengths) were lower in their inherent sensitivity than the longer wavelength detectors (Carlson et al. 1992). We use the average of the temperatures and areas derived from each grating as the measured temperature and area and the standard deviation in the measurements as the uncertainty. The left panel of figure 2 shows the resulting blackbody fit superimposed on the measured brightness of a hotspot observed in observation G1INNSPEC_01A at 0 N 78 W, likely Hi'iaka. The plus signs are the total brightness of the hotspot while the solid line is a blackbody curve for a temperature of 439 K and an area of 57 km$^2$, the average of the temperatures and areas found for each of the grating positions individually. We only used data points at wavelengths longer than 3.5 µm, indicated by the vertical line, in our fit.

For observations obtained in only a single grating position, we do not calculate formal uncertainties, but we expect the results to be uncertain by a factor of a few. We note the temperature and emitting area (in km$^2$) of the blackbody curve. We use the 3.5-µm brightness of the resulting curve fit as an estimate of the strength of that volcanoes emission.

**Daytime Data**

We estimate the reflected sunlight contribution at each pixel by assuming that the 1 - 3 micron signal is dominated by sunlight, and also that the reflectance is independent of wavelength. Using these assumptions, we scale a solar spectrum to match the 1 - 3 micron NIMS spectrum for the pixel, and subtract the scaled solar spectrum from NIMS spectrum over the full wavelength range. This method assumes that there is negligible contribution from volcanic emission between 1 and 3 µm. It also does not take into account the strong $SO_2$ absorption bands. Because of this, we use only the parts of the spectrum away from the major absorption bands, namely 3.0-3.2 μm and 4.8-5.0 μm. Because we only used data points from these two narrow wavelength bands in determining a single temperature blackbody fit, we only considered images that had many wavelengths so that there was more than a single data point in each of these bands. Furthermore, we were unable to derive formal uncertainties, but we expect the results to be uncertain by a factor of a few. The right panel of figure 2 shows the resulting blackbody fit superimposed on the measured brightness of a hotspot observed in observation G2INHRSPEC01A at 18 N 174 W, likely Pillan. The solid line is a blackbody curve for a temperature of 392 K and an area of 751 km$^2$. The vertical lines indicate the two narrow wavelength bands that were used for the fit.

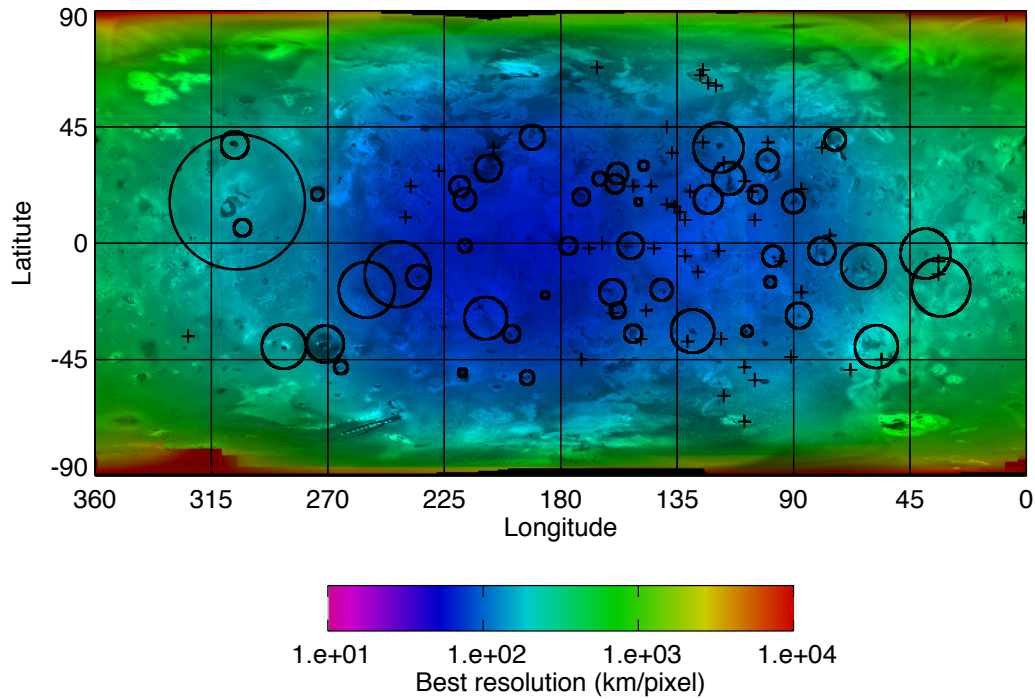

Figure 3: The colors on this map indicate the best resolution of all the available NIMS cubes for that location. The resolution was calculated in the N-S direction and emission angle is accounted for. We consider the area between 70 W – 280 W and 70 S - 70 N to be well covered by NIMS for hotspot detection. Circles, plus signs, and background map are the same NIMS data from figure 1.

     Using the systematic process described above, we analyzed over 100 NIMS observations (both nighttime and daytime) obtained between June 1996 and May 1999 (prior to the close flyby observations which are not used here as they cover smaller regions of Io and were acquired using fewer wavelengths) and made 287 measurements of active hotspots (see appendix). The hotspots included 51 distinct volcanic features (figure 1). Note that not all hotspots detected by NIMS are included in this study, for example, we did not include the regional-level, higher spatial resolution observations obtained during Galileo flybys of Io. Several volcanoes were observed by NIMS for which we were unable to obtain blackbody fits due to low intensity or noisy data. However, these hot spots were summarized in Appendix 1 of Lopes and Spencer

(2007). We include in figures 1 and 3 hotspots listed as being observed by NIMS that we were unable to measure brightnesses for, but we do not include them in our subsequent analysis.

Of the 287 measurements we made of active hotspots, we were able to determine formal uncertainties for 81 (28%), the remainder were obtained in a single grating position or during daytime observations and, as a result, could be uncertain by a factor of a few. Two of the measurements had unreasonably high formal uncertainties of greater than a factor of 10. The average uncertainty of the remaining 79 measurements with formal uncertainties was 74% with a median of 41%. Of the 51 distinct volcanic features, half had at least one measurement with formal uncertainties and nearly two thirds had multiple measurements with an average standard deviation of 85%. As such, the NIMS observations used in our analysis are certain to at least a factor of 2. Since the brightnesses of these volcanoes varies by more than 4 orders of magnitude, the factor of 2 is negligible.

In order to use the brightnesses of the 51 volcanic features as a proxy for heat flow, we take the average of all determined 3.5-$\mu$m brightnesses for each volcano to determine a single value for each volcano. While NIMS obtained global coverage, it obtained significantly higher resolution images of the anti-Jupiter hemisphere. Figure 3 shows, for each point on the surface, the number of observations and best resolutions available from the NIMS data we used. We account for emission angle in the NS direction by dividing the resolution at the subspacecraft point by the cosine of the emission angle. Note that this plot was made using the spatial resolution of the NIMS cubes, which are oversampled by a factor of 2 from the original observations. The identified hotspots are generally located in areas where the best available image resolution is better. The average image resolution available is approximately 300 km/pixel (for a NIMS cube, which is oversampled), while most hotspots were found in areas with images resolutions better than 150 km/pixel. Areas on Io with NIMS cube resolutions better than about 150 km/pixel are equatorward of 70 degrees latitude and between longitudes of 70 and 280 W. This is the area we consider well covered for NIMS measurements of active volcanic activity.

## Comparison to Tidal heating models

### Description of Tidal heating models

While it is well known that Io's volcanism is due to tidal dissipation in its interior, how this process operates is not understood. Two end-member models have been proposed for the distribution of the tidally generated heat dissipation in Io and each has implications for the distribution of upwellings and downwellings (Tackley, et al., 2001; Segatz et al., 1988; Ross et al., 1990; Gaskell et al, 1988). Both of these end-member models assume that convection is the main mode of heat transfer within Io, but differ on where the tidal heating predominantly occurs, and, therefore, the nature of Io's interior. A reasonable assumption is that the distribution of hotspots on the surface mimics the distribution of convective upwellings and thus can be used to differentiate between the models. If the heating occurs mainly in the deep mantle, upwellings occur at the poles with downwelling at the equator, leading to 2-fold symmetry. If the heating occurs mainly in the asthenosphere, the major downwelling occurs at the poles with two major and two minor upwellings around the equator, resulting in upwelling separations of a few hundred kilometers. Thus, measurements of the spatial variability of Io's heat flow can distinguish between these models, or test more complex models that involve components of both, such as the model of Tackley et al. (2001).

The modeling of Io's tidal heating by Tackley et al. (2001) has found that the large-scale mantle convective flow pattern is dominated by the distribution of tidal heating (mantle or asthenosphere) with smaller heat flux variations as the Rayleigh number increases. They also found that the mixed heating model of 2/3 asthenosphere and 1/3 mantle favored by Ross et al. (1990) results in a surface expression with 4-fold symmetry of upwellings as opposed to the 2-fold found in the mantle case and upwellings focused in lobes at 45 degrees latitude as opposed to the equator as in the asthenosphere case. They agreed with earlier conclusions that the observed distribution of volcanic centers known at that time is best fit by the asthenosphere model. More recent heat flow modeling by Hamilton et al. (2013) roughly agrees with the results of Tackley et al. (2001) and Ross et al. (1990) in that the mantle heating model predicts higher heat flow at the poles and lower at the equator, with minima at the sub- and anti-Jovian points and that the asthenospheric heating model predicts lower heat flow at the poles and higher near the equator, with maxima at 90° intervals (longitudes 0°, 90°, 180°, and 270°).

Tyler et al. (2015) considered tidal heating of Io including a fluid magma ocean. The results of their modeling are qualitatively similar to the solid asthenospheric heating end member in that it exhibits 2-fold symmetry and the majority of heat flux occurs near the equator. However, because of the fluid layer, the peaks in heating are shifted approximately 30 degrees to the east. They find that this is a better match with the calculated locations of volcano clusters on Io (Hamilton et al., 2013).

### Comparison of Models to Data

We compare the heat flow predictions of the asthenospheric heating and mantle heating models of Ross et al. (1990) and Hamilton et al. (2013) to our derived volcanic power outputs using 2 methods to aggregate the hot spot data: modified counting circles and histograms.

### Modified Counting circles

Tackley et al. (2001) used standard counting circles of 15° in radius to illustrate the anti-correlation between Io's global distribution of volcanoes and mountains. We use a similar method to map the heat flow from the volcanic power output. For each data set, we create a map with a resolution of 10° in both latitude and longitude. For each point in the map, we determine the total power output of all volcanoes within 30° of the center of the bin and divide by the surface area of a 30° spherical cap (a portion of a sphere cut off by a plane), to calculate the heat flow or brightness density in that bin. We experimented with bins of smaller sizes, but with the resolution of the data sets and sparsity of volcanoes with measured powers, the 30° bins yielded the maps that were easiest to interpret. In the PPR data, Loki is nearly an order of magnitude more powerful than the next most powerful volcano, so we show the PPR data both with and without Loki. Figure 4 shows the resulting maps for the volcanoes measured from Galileo PPR, Galileo NIMS, and New Horizons LEISA.

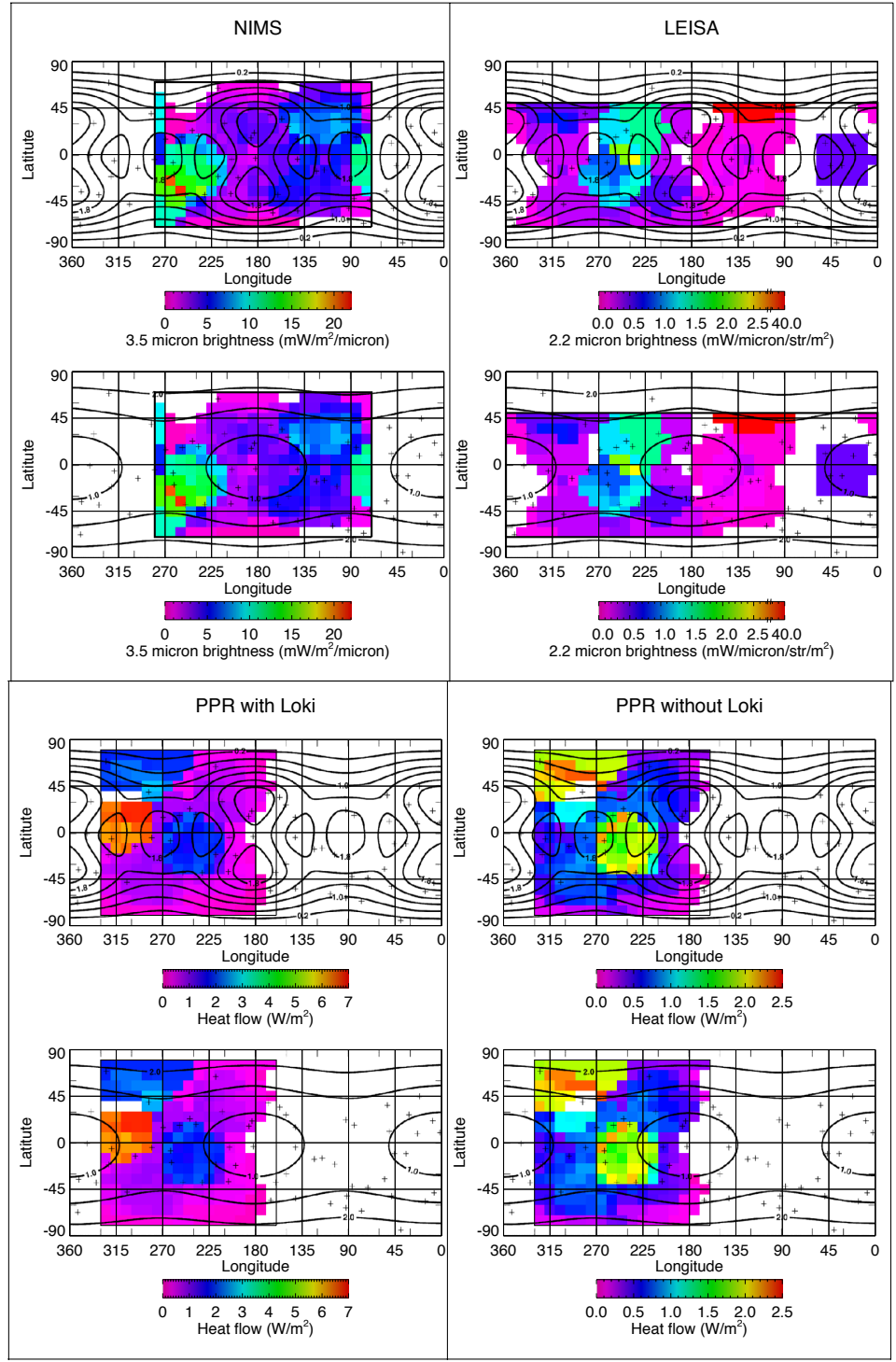

**Figure 4: Maps of brightness (or power) density using 30° counting circles. The lower panels show PPR data, upper left is NIMS, and upper right is LEISA. In all cases the contours show the predicted heat flow from Ross et al. (1990) with the top plot for asthenospheric heating and the bottom for mantle heating. The faded area outside the boxes for PPR and NIMS indicate areas without good coverage.**

One of the most obvious differences between the asthenospheric and mantle heating models of Ross et al. (1990) occurs at high latitudes where the heat flow is predicted to be at a maximum in the mantle heating model and a minimum in the asthenospheric model (figure 4). With the exception of Tvashtar, observed in the LEISA data, few volcanoes have been observed at high latitudes. However, this may be due to observational biases because of the large emission angles at high latitudes, although hot spots have been easily observed near the limb at lower latitudes (Lopes-Gautier, et al., 1999).

If we consider only the lower latitude data to distinguish between models, another obvious difference occurs near the equator. The mantle heating model predicts a dearth of heat flow at the sub- and anti-Jovian points (longitudes 0° and 180°) while the asthenospheric model predicts an increase in heat flow at the sub- and anti-Jovian points as well as the center of the leading and trailing hemispheres (longitudes 90° and 270°). All three data sets show a dearth of volcanoes near the sub- and anti-Jovian points, suggesting they are more compatible with the mantle-heating model. However, there are some observational biases, as NIMS had poor coverage of the sub-Jupiter point and PPR had poor coverage of the anti-Jupiter point.

**Histograms**

Hamilton et al. (2013) analyzed the spatial variation in volcanic centers by plotting a histogram of the number of volcanoes per unit area by latitude and longitude. They found that the number of volcanoes per unit area was greatest near the equator and dropped towards the poles and was greatest at the sub- and anti-Jovian points, decreasing towards the leading and trailing hemispheres, more consistent with the asthenospheric heating model.

We make similar histograms, but instead of plotting number of volcanoes per unit area, since we have measurements of power and brightness, we plot the total power or brightness of all volcanoes per unit area for each latitude and longitude bin, using data from NIMS, LEISA, and PPR. Our results are shown in figure 5. All of the data sets show reduced power and brightness density near the poles, but they also show a decrease near the equator. Veeder et al. (2012) also found that the power output of volcanoes, per unit surface area, is slightly biased towards mid-latitudes. If we again ignore the behavior near the poles, the observed latitudinal variation is more consistent with the mantle-heating model.

All three of our data sets also show a dearth of heat flow at the sub- and anti-Jovian points. Davies et al. (2015), Veeder at al. (2012), and de Kleer and de Pater (2016) also show a similar drop in their calculated power outputs near these locations. The locations of these minima are more consistent with the mantle-heating model.

The LEISA data show four-fold symmetry in longitude, with deficiencies at 0°, 90°, 180° and 270° and peaks near 45°, 135°, 225°, and 315°. While the PPR and NIMS data do not cover the complete longitude range, they are consistent with both the 4-fold symmetry observed in the LEISA data and 2-fold symmetry. Four-fold symmetry is consistent with asthenospheric heating than mantle, but is even more consistent with the combined heating case of Tackley et al. (2001). Additionally, this model predicts a peak in heat flow near 45° N and S instead of the equator. This is also consistent with our measurements from all three instruments in that they all show a dearth of power at the equator and peaks at mid-latitudes. The asthenospheric heating model has peaks at 0°, 90°, 180° and 270°, so an eastward shift of 45° in the LEISA data would result in a match with this model. Previous studies also suggested a 30° - 60° eastward shift (Hamilton et al. 2013; Kirchoff et al., 2011; de Kleer and de Pater, 2016). However, the asthenospheric model also predicts a peak heat flow near the equator, while all the data examined here show a dearth in heat flow at the equator and a peak at low to mid latitudes.

**What's going on at high-latitudes?**

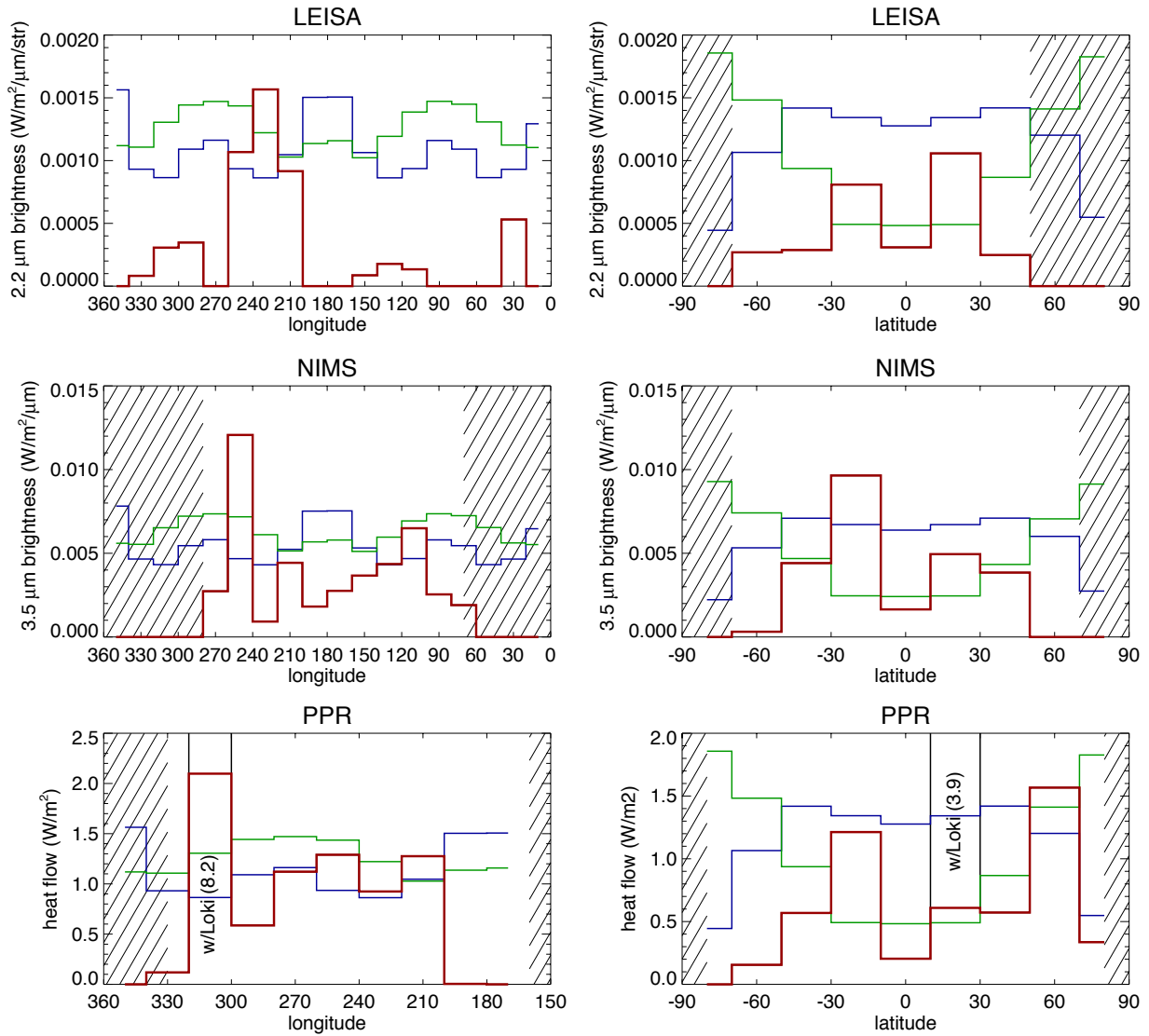

Figure 5: Histograms of power and brightness per unit area for PPR (bottom), NIMS (center), and LEISA (top). The left column shows variations with longitude and the right column variations with latitude. The red and black lines are the measured data. The blue and green lines show histograms of the heat flow from the asthenospheric model (blue) and mantle model (green) from Hamilton et al. (2013), in arbitrary units. Hatching indicates areas that are not well covered by each instrument.

    Beyond the fact that hot spots are not a perfect approximation of heat flow, one of the largest difficulties in comparing observations of volcanic eruptions with heat flow models is the observational biases in the observations, especially at high latitudes. This is particularly important because the largest difference between the two end member heat flow models occurs at high latitudes. Of the three data sets analyzed, the NIMS data set included the fewest volcanoes at high latitudes (greater than 60°), none, while 5.8% of the volcanoes observed by PPR and 2.7% by LEISA were at high latitudes. Furthermore, approximately half the volcanoes found by PPR and LEISA were at low latitudes (less than 30°), while for NIMS the fraction is 71%. Half

of Io's surface area is at low latitudes, so the PPR and LEISA results are consistent with a uniform distribution. This is not a simple wavelength dependence, since NIMS observations occurred at the middle of the wavelength range. The easiest explanation would be that NIMS had the most difficulties observing volcanoes at high emission angles, but Lopes-Gautier et al. (1999) found that this was not the case, as volcanoes near the limb were observed. All three datasets have observational biases that favored low latitude observations. All had sub-spacecraft points within 5° of the equator, with PPR and NIMS closest to the equator and LEISA between 5 and 10 S, which lead to large emission angles at high latitudes. Recent ground-based observations obtained at wavelengths similar to NIMS detected only one volcano (2.1 %) at high latitudes and did not detect any volcanoes polewards of 65° (de Kleer and de Pater, 2016). If eruptions were randomly distributed on Io's surface, 9% of eruptions should occur at these high latitudes.

Rathbun et al. (2004) found that most PPR hotspots were point sources in the PPR data. The exceptions were a few high-latitude hotpots which were large enough to be spatially resolved, consistent with the hypothesis that high-latitude eruptions occur less frequently, but at larger scales. The only volcano observed by LEISA at high latitude was Tvashtar, which was not measured either our PPR or NIMS global data sets, suggesting that New Horizons may have witnessed a rare, large, high-latitude eruption. The fact that Tvashtar was the brightest eruption observed by LEISA, by more than order of magnitude, is consistent with the hypothesis that high-latitude eruptions are rarer and more powerful (Milazzo et al., 2005; Rathbun et al., 2004; Lopes et al., 2004; Marchis et al., 2002; Radebaugh et al., 2001; Howell et al., 2001). If this is the case, our current data sets are unable to capture the long-term heat flow from high-latitude volcanoes. The one exception may be Galileo PPR observations, which, due to their longer wavelength, are able to detect lava flows hundreds of years old and thus potentially measure heat flow integrated over timescales longer than even the high-latitude eruption frequency. However, Tvashtar's heat flow is concentrated in discrete lava lakes (Milazzo et al., 2005; Rathbun et al., 2014) and not in large lava flows like other high latitude hotspots. de Kleer and de Pater (2016) found that large eruptions occur at higher latitudes and that persistent hotspots occur at lower latitudes, consistent with our results. Davies et al. (2015) suggested that the warm poles observed by Galileo PPR (Rathbun et al., 2004) may be the result of heat flow from deep-mantle heating.

In order to gain further insight into the behavior of Io's heat flow at high latitudes, we took a closer look at the Galileo PPR observations, particularly the surprisingly constant nighttime temperatures. Rathbun et al. (2004) show that the diurnal temperatures are remarkably constant at night, decreasing little towards morning or towards the poles (see figs. 3 and 7 in Rathbun et al., 2004). In order to match the diurnal variation in nighttime temperatures within 20 degrees of the equator, Rathbun et al. (2004) required an inhomogeneous surface composed of two units: a high albedo high thermal inertia unit plausibly representing surface $SO_2$ frost, and a low albedo low thermal inertia unit. Here, we model the nighttime surface temperature from an equator to south pole obtained by PPR during Galileo's 31$^{st}$ orbit (figure 6). For each observation, we determine the bolometric albedo at that location using the map derived from Galileo images by Simonelli et al. (2001). We fix the albedo for the low albedo unit to be 0.34 (Rathbun et al., 2004) and adjust the albedo of the high albedo unit so that a 50/50 ratio of both units gives the average equatorial albedo found by Simonelli et al. (2001). The average albedo is 0.534, so our high albedo unit has an albedo of 0.656. We find the best fit at low latitudes has a thermal inertia of $3.0 \times 10^4$ $J/m^2/K/s^{1/2}$ for the low albedo unit and $5.0 \times 10^5$ $J/m^2/K/s^{1/2}$ for the high albedo unit. We determine the areal fraction of each unit at each location using the albedos from Simonelli et al. (2001). This model cannot explain the surprisingly high polar temperatures: poleward of about 35°, the measured temperature is substantially higher than the predicted temperature.

An additional complication to this analysis is that PPR maps of nighttime temperature show that the outer reddish-orange ring of Pele's plume deposits is consistently 5 K higher than the surrounding terrain (Rathbun et al., 2004). This PPR temperature scan intersects the ring between latitudes -30 and -45, corresponding to a temperature increase in the PPR data. Further, temperatures measured at high-latitude could be disproportionally observing equator-facing slopes. Thermal modeling of Mars found that temperatures could be increased up to 9 K for equator-facing slopes of 0.1 radian at latitudes lower than 70 (Kieffer, 2013). Other possible explanations of the high nighttime temperatures at high latitude include additional endogenic heating (Veeder et al., 2004) or small hotspots below the resolution of the various instruments.

To truly understand the spatial variation in Io's heat flow, better thermal observations of Io's poles are required.

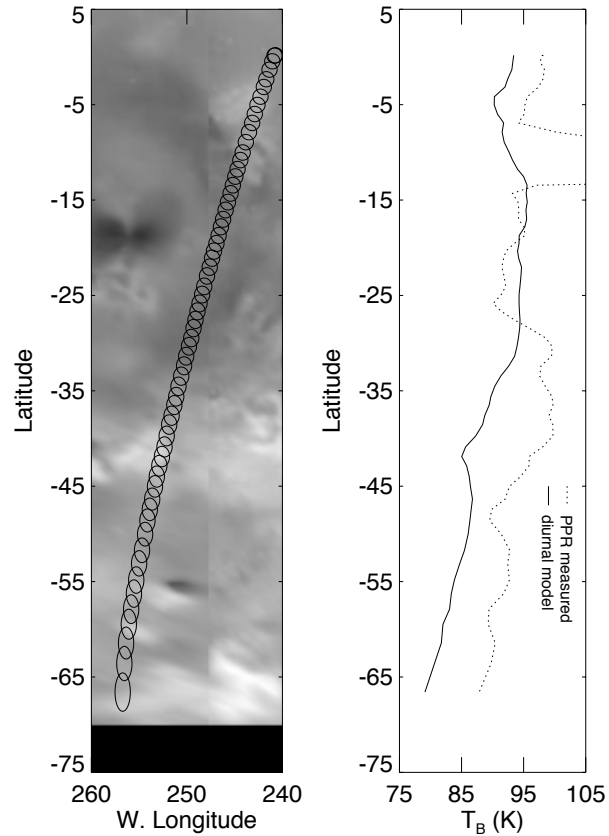

**Figure 6: Left panel shows location of PPR fields of view overlaid on SSI image mosaic. Right panel shows the measured (dotted line) and modeled (solid line) temperatures. Except for where the observations cross an obvious volcano, the model matches the data equator ward of ~30 S.**

With those observations, we can eliminate biases due to emission angle and surface slopes. Further, consistent monitoring of Io's volcanoes will enable us to determine if high-latitude eruptions are rarer and more powerful.

**Conclusions**

We compared observations of active volcanoes from Galileo PPR, NIMS, and New Horizons LEISA to tidal heating models. This comparison assumes that the areal density of volcanic output is a good proxy for heat flow. None of the data available are able to conclusively distinguish between the end member tidal heating models. High latitude observations may suffer from observational biases and other differences between the models are small. However, all three data sets show a relative dearth of volcanic brightness near 180 W (anti-Jovian point) and near the equator. Furthermore, the only data set with good observations of the sub-Jovian point (LEISA) also shows a lack of volcanic brightness there. Davies et al.

(2015) and de Kleer and de Pater (2016) also found a dearth of volcanic brightness at the sub- and anti-Jupiter points.

Our observations are more consistent with the mantle-heating model than the asthenospheric-heating model. However, the mantle and asthenospheric heating models are end-members of a range of possibilities. All of our data sets are consistent with four-fold symmetry, which matches the combined heating case of Tackley et al. (2001), supporting this combined heating model. The dearth of volcanoes observed near the equator is also consistent with this combined model. Of course, the observational biases at high latitudes make it difficult to determine between mantle and asthenospheric heating. de Kleer and de Pater (2016) found that bright eruptions and hotspots on the trailing hemisphere correspond better to the mantle-heating model while hotspots in the leading hemisphere and persistent hotspots correspond better to recent models incorporating a partially-fluid interior. Davies et al. (2015) finds that a ratio of deep to shallow heating of 1:1 is the best match to their results, which include Io's anomalously warm poles.

**Acknowledgements**

We would like to thank Fabíola Magalhães and Sebastian Saballett who assisted with some of the NIMS data reduction and anonymous reviewers for improving the paper. Part of this work was conducted at Jet Propulsion Laboratory, California Institute of Technology, under contract with NASA. This project was supported by the NASA Outer Planets Research Program.

Appendix

Table 1: Brightness of hotspots measured in NIMS observations by the systematic process described.

| Date of observation | | | | | | | 3.5-μm | |
| Year | Month | day | Observation | Latitude | Longitude | Name of volcano | flux (W/μm) | Uncertainty in flux |
|---|---|---|---|---|---|---|---|---|
| 1996 | 6 | 28 | G1INCHEMIS01A | -28.4 | 88.5 | Ekhi | 2.43E+09 | |
| 1996 | 6 | 28 | G1INCHEMIS01A | 32.5 | 100.5 | Shango | 1.78E+09 | |
| 1996 | 6 | 28 | G1INCHEMIS01A | 19.8 | 104.8 | Monan | 1.28E+09 | |
| 1996 | 6 | 28 | G1INCHEMIS01A | -9.9 | 63.2 | Shamshu | 7.53E+09 | |
| 1996 | 6 | 28 | G1INCHEMIS01A | -40 | 58.4 | Laki-Oi | 6.91E+09 | |
| 1996 | 6 | 28 | G1INNSPEC_01A | 25 | 115.2 | Amirani | 2.39E+09 | 2.62E+08 |
| 1996 | 6 | 29 | G1INNSPEC_01A | -34.4 | 108.8 | Altjirra | 7.91E+08 | 1.98E+08 |
| 1996 | 6 | 30 | G1INNSPEC_01A | -34.2 | 129.7 | Malik | 1.31E+09 | 4.03E+08 |

Remainder of the table is available in actual paper.